\documentclass{svjour3}       
\smartqed  
\usepackage{graphicx}
\journalname{General Relativity and Gravitation}
\newcommand{\gtorder}{\mathrel{\raise.3ex\hbox{$>$}\mkern-14mu
            \lower0.6ex\hbox{$\sim$}}}
\newcommand{\ltorder}{\mathrel{\raise.3ex\hbox{$<$}\mkern-14mu
            \lower0.6ex\hbox{$\sim$}}}

\begin{document}

\title{Implications of the Gravitational Wave Event GW150914}

\author{M. Coleman Miller
}


\institute{Department of Astronomy and Joint Space-Science Institute, University of Maryland, College Park, MD 20742-2421
              \email{miller@astro.umd.edu}}

\date{Received: date / Revised: date}

\maketitle

\begin{abstract}
The era of gravitational-wave astronomy began on 14 September 2015, when the LIGO Scientific Collaboration detected the merger of two $\sim 30~M_\odot$ black holes at a distance of $\sim 400$~Mpc.  This event has facilitated qualitatively new tests of gravitational theories, and has also produced exciting information about the astrophysical origin of black hole binaries.  In this review we discuss the implications of this event for gravitational physics and astrophysics, as well as the expectations for future detections.  In brief: (1)~because the spins of the black holes could not be measured accurately and because mergers are not well calculated for modified theories of gravity, the current analysis of GW150914 does not place strong constraints on gravity variants that change only the generation of gravitational waves, but (2)~it does strongly constrain alterations of the propagation of gravitational waves and alternatives to black holes.  Finally, (3)~many astrophysical models for the origin of heavy black hole binaries such as the GW150914 system are in play, but a reasonably robust conclusion that was reached even prior to the detection is that the environment of such systems needs to have a relatively low abundance of elements heavier than helium.
\end{abstract}

\section{Introduction}
\label{sec:intro}
The long-awaited first direct detection of gravitational waves occurred at 09:50:45 UTC on 14 September 2015, shortly {\it before} the official start to the O1 science run of Advanced LIGO: the Advanced Laser Interferometric Gravitational-wave Observatory \cite{2016PhRvL.116f1102A}.  This represents the culmination of a century-long effort that started with Einstein's original descriptions of gravitational waves \cite{1916SPAW.......688E,1918SPAW.......154E} and continued with the demonstration that close binaries with two neutron stars spiral together at the rate predicted by general relativity \cite{1975ApJ...195L..51H,1979Natur.277..437T,1982ApJ...253..908T}.  It also represents the beginning of what is expected to be an exploration of a new window to the universe, through which we can see extreme and previously invisible events. 

Here we present an astrophysicist's view of some of the implications of this event for gravitational physics and astrophysics, as well as the prospects for much more information in the future.  In \S2 we start with general background and then review the event itself: the properties of the detector and the detection and the checks made on it.  We note that there was a less-significant candidate on 12 October 2015, called LVT151012, that was probably another double black hole coalescence, but in the spirit of conservatism we will largely ignore this possibility because additional strong events are expected to be seen in later data.  In \S3 we discuss the importance of the detection for tests of gravitational theories and of exotic alternatives to black holes.  In \S4 we turn our attention to the implications of the event for the astrophysics of black holes and the variety of formation mechanisms that have been proposed for GW150914.  We conclude in \S5 by discussing briefly the expectations for information from future events, including the prospects for learning about cold dense matter in neutron stars, localization of events and the possibility of electromagnetic counterparts, and the prospects for gravitational wave detection in other frequency bands.  

\section{Gravitational Waves, the Detector, and the Detection}
\label{sec:gw}

In this section we begin by reviewing some of the basic aspects of gravitational waves.  We then discuss the Advanced LIGO detector and the main GW150914 detection.

\subsection{Basics of gravitational waves}
\label{sec:gwbasics}

Radiation (whether gravitational, electromagnetic, or other) requires time variation of the source.  We can therefore get insight into what kinds of sources can produce gravitational radiation by breaking a source into moments and determining which of those moments can vary.

Let the mass-energy density be $\rho({\bf r})$.  The monopole moment we choose is $\int\rho({\bf r})d^3r$, which is simply the total mass-energy of the source.  This is constant, so there cannot be monopolar gravitational  radiation.  The static dipole moment is $\int\rho({\bf r}){\bf r}d^3r$.  Conservation of linear momentum guarantees that a comoving observer will not see this moment change, so there cannot be dipolar gravitational radiation.  The dipolar current is $\int\rho({\bf r}){\bf r}\times{\bf v}({\bf r})d^3r$.  This, however, is simply the total angular momentum of the system, so its conservation means that gravitational radiation is not produced at this order.  The next moment is quadrupolar: $I_{ij}=\int\rho({\bf r})r_ir_j d^3r$.  This moment need not be conserved, so there can be quadrupolar gravitational radiation.

These considerations allow us to draw general conclusions about the types of astronomical sources that can generate gravitational radiation.  A spherically symmetric variation is only monopolar, and thus it does not produce radiation.  No matter how violent an explosion or a collapse, no gravitational radiation is emitted if spherical symmetry is maintained.  In addition, a rotation that preserves axisymmetry (without contraction or expansion) does not generate gravitational radiation because the quadrupolar and higher moments are unaltered.  Therefore, for example, a neutron star can rotate arbitrarily rapidly without emitting gravitational radiation as long as it maintains axisymmetry and stationarity.

This immediately allows us to focus on the most promising types of sources for gravitational wave emission.  The general categories are: binaries, continuous wave sources (e.g., rotating stars with nonaxisymmetric lumps), bursts (e.g., asymmetric collapses), and stochastic sources (i.e., the superposition of many individually unresolvable sources; a particularly interesting example in this category would be a background of gravitational waves from the early universe, but another example is the foreground from the large population of double white dwarf binaries in our Galaxy).

We now focus on binary systems.  The characteristics of the gravitational waves from binaries, and what we could learn from them, depend on the nature of the objects in those binaries.  The typical dimensionless gravitational wave strain (i.e., the fractional amount by which a separation between test masses changes as a wave goes by) measured a distance $r$ from a well-separated circular binary is \cite{1997rggr.conf..447S}
\begin{equation}
h=2(4\pi)^{1/3}{G^{5/3}\over c^4}f_{\rm GW}^{2/3}{\cal M}^{5/3}
{1\over r}\; ,
\end{equation}
where $f_{\rm GW}$ is the gravitational wave frequency and ${\cal M}=\mu^{3/5} M^{2/5}$ is the ``chirp mass", where $M=m_1+m_2$ is the total mass of a binary with component gravitational masses $m_1$ and $m_2$ and $\mu=m_1m_2/M$ is the reduced mass.  This formula does not include redshifts.

If the binary orbit is eccentric, then to lowest order in an expansion in small velocities the evolution of the semimajor axis $a$ and eccentricity $e$ of the orbit is given by the formulae \cite{1963PhRv..131..435P,1964PhRv..136.1224P}
\begin{equation}
\biggl\langle{da\over{dt}}\biggr\rangle=-{64\over 5}{G^3\mu M^2\over{
c^5a^3(1-e^2)^{7/2}}}\left(1+{73\over{24}}e^2+{37\over{96}}e^4\right)
\label{eq:aPeters}
\end{equation}
and
\begin{equation}
\biggl\langle{de\over{dt}}\biggr\rangle=-{304\over{15}}e{G^3\mu M^2\over{
c^5a^4(1-e^2)^{5/2}}}\left(1+{121\over{304}}e^2\right)
\label{eq:ePeters}
\end{equation}
where the angle brackets indicate an average over an orbit.  One can
show that these formulae imply that the quantity
\begin{equation}
ae^{-12/19}(1-e^2)\left(1+{121\over{304}}e^2\right)^{-870/2299}
\end{equation}
is constant throughout the inspiral \cite{1964PhRv..136.1224P}, and that the time to coalesce for a circular binary of initial semimajor axis $a_0$ is \cite{1964PhRv..136.1224P}
\begin{equation}
T_{\rm GW}(e=0)\approx 6\times 10^8~{\rm yr}~(a_0/0.1~{\rm AU})^4~(\mu/15~M_\odot)^{-1}~(M/60~M_\odot)^{-2}\; .
\end{equation}
For an eccentric binary of initial eccentricity $e$, the coalescence time is $T_{\rm GW}(e)\approx (768/425)(1-e^2)^{7/2}T_{\rm GW}(e=0)$ \cite{1964PhRv..136.1224P}.

For high eccentricity, $1-e\ll 1$, the pericenter distance $a(1-e)$ is approximately conserved while the apocenter distance $a(1+e)$ decreases.  For low eccentricity $e\ll 1$, $ae^{-12/19}$ is approximately conserved, which means that the eccentricity decreases with the orbital frequency $f\propto a^{-3/2}$ like $e\propto f^{-19/18}\sim f^{-1}$.  Thus as the orbit shrinks due to gravitational radiation, the orbit circularizes (until the objects get very close to each other).  The large sizes of typical binary orbits ($\gtorder 10^{11}$~cm) compared with the small sizes of neutron stars and stellar-mass black holes ($\ltorder 10^7$~cm) leads to the strong expectation that most stellar-mass compact binaries in the Advanced LIGO frequency range will be close to circular (but see \S~\ref{sec:astro} for further discussion).

In general relativity (GR), gravitational disturbances travel at the speed of light.  This is equivalent to the statement that in a quantum theory of gravity the carrier of the gravitational interaction (called the graviton) is massless.  This implies that, like light in a vacuum, gravity propagates as a transverse wave that has two polarizations.  The effect of these polarizations on a ring of test particles whose axis is along the direction of propagation of the wave is to change their separations in characteristic ``+" and ``$\times$" patterns.  The strain can thus be broken into a ``+" component $h_+$ and a ``$\times$" component $h_\times$.  In other theories of gravity, different polarization modes are possible.

Prior to GW150914, the best tests of these ideas came from careful timing of pulsars in double neutron star binaries.  Pulsars are spectacularly precise clocks, so even small variations in their timing can be seen (see \cite{2009arXiv0907.3219F} for a review).  When a pulsar is in a binary with another neutron star, the pointlike nature of both objects means that the systems are free of interference from most astrophysical complications.  Thus many subtle effects, including the inspiral due to emission of gravitational radiation, can be measured.  The rate of reduction of the orbital period is consistent with the predictions of general relativity at the 0.1\% level for several sources \cite{2009ASSL..359...43K}.  But no clean tests of the predictions of gravitational theories in extreme gravity were available prior to the direct detection of gravitational waves.  Electromagnetic observations provide a valuable counterpart to gravitational wave detections, but the complexities involved in the electromagnetic interactions of matter make it more difficult to draw direct conclusions.

\subsection{Gravitational-wave detectors}
\label{sec:detectors}

The effect of gravitational waves as they pass through a detector is to change the separations and relative rates of time flow between different parts of the detector.  We will focus here on ground-based detectors. However, we note that pulsar timing arrays, which aim to discover gravitational waves with frequencies $\sim 10^{-9}-10^{-6}$~Hz via precise timing of multiple pulsars, are also rapidly increasing in sensitivity \cite{2016MNRAS.458.1267V}, and that space-based detectors such as eLISA are planned to fly in the next two decades \cite{2013arXiv1305.5720C}.

Ground-based measurements fall in two categories.  The first involves resonant detectors, usually bars or spheres \cite{1960PhRv..117..306W,2011RAA....11....1A}.  The idea is that a gravitational wave will induce oscillations in the detector, and if the frequency of the wave is close enough to the resonant frequency and the device has a high enough quality factor, then those oscillations can then be measured by, e.g., piezoelectric detectors.  Such detectors have some advantages (e.g., spherical detectors can measure many polarization modes at once \cite{1992mgm..conf..988H}), but they tend to be sensitive only over a restricted range of frequencies near the resonant frequency.  For reasons of cost the detectors have to be relatively small, with the consequence that the resonant frequencies are typically $\sim 3000$~Hz.  This is above the expected frequency during inspiral but may be close to that of the f-mode of a merged double neutron star remnant \cite{2012PhRvD..86f3001B} or oscillations in supernovae or giant flares from soft gamma-ray repeaters \cite{2013CQGra..30l3001B}.  Even at the resonant frequency, these detectors tend not to be as sensitive as the km-scale laser interferometers.  For example, the expected noise level of the MiniGRAIL spherical detector at its 2942.9~Hz resonant frequency will be $\approx 10^{-21}$~Hz$^{-1/2}$ when it operates at 50~mK, and a quantum limited detector of this kind would reach $\approx 10^{-22}$~Hz$^{-1/2}$ \cite{2007PhRvD..76j2005G}.  In contrast, at its design sensitivity Advanced LIGO is expected to reach noise levels of $\approx 10^{-23}$~Hz$^{-1/2}$ at 3000~Hz \cite{2016arXiv160400439T}.  We therefore turn our attention to km-scale laser interferometers.

The idea of using laser interferometry to detect gravitational waves was first proposed by \cite{1963JETP...16..433G}, and was developed further in many subsequent papers (see \cite{1971ApOpt..10.2495M,1983grg1.conf..887D,1992Sci...256..325A} for some representative examples).  A classic discussion of the noise sources for such an interferometer is given by Weiss \cite{1972QPRRLE..105.54W}; remarkably, even his estimates for the magnitudes of the noise sources hold up well today (see \cite{2015CQGra..32k5012A} for a recent characterization of the detectors).  In the LIGO detectors, as well as in GEO600, Virgo (currently being upgraded), KAGRA (currently being constructed), and the planned LIGO-India detector (see \cite{2015JPhCS.610a2012D} for a recent discussion of the status of several of these detectors; note that final approval for LIGO-India was given just a week after the announcement of the GW150914 event), the instrument is L-shaped, with equal arm lengths that range from 600 meters (GEO600) to 4~km (LIGO and LIGO-India).  Effectively, laser light is bounced back and forth many times between the corner station and each of the end stations (this is made possible by the introduction of an extra set of mirrors between the beam splitter and the end test masses, which forms a pair of resonant Fabry-Perot cavities), and interference patterns are obtained from the light returning from the end stations.  There are many improvements to the Advanced LIGO detector compared with its earlier incarnations, including greater laser power, better mirrors and coatings, and better suspensions, and still more are to come.  The full design-sensitivity Advanced LIGO will be able to detect double neutron star mergers out to nearly three times farther than would have been possible in the first Advanced LIGO observing run (called the O1 run; see \cite{2016PhRvL.116m1103A}), and the effective sensitivity will increase by a factor $\gtorder 10$ for heavy black holes \cite{2016LRR....19....1A}.  Because gravitational wave detectors sense amplitude, not flux, a sensitivity increase by some factor increases the volume that can be probed by the cube of that factor (if redshifts are small).  There is every reason to believe that GW150914 is only the beginning.

\subsection{The 14 September 2015 event}
\label{sec:event}

The signal from GW150914 entered the Advanced LIGO band at $\sim 35$~Hz.  Over the next $\sim 0.2$~seconds the gravitational waves from this event swept up in frequency and amplitude, then rung down.  The signal was detected first using a burst analysis, i.e., an analysis that simply looked for correlated excess signal beyond the inferred noise level without any presumption that the event was a binary merger \cite{2008CQGra..25k4029K,2015CQGra..32m5012C,2016arXiv160203843T}.  Later analysis in the context of compact binary coalescence determined that the merger was of two black holes, both with masses $\sim 30~M_\odot$, without any strong evidence for spin or eccentricity \cite{2016PhRvL.116f1102A,2016arXiv160203840T}.  

In the majority of this review we will discuss the implications of this event for gravitational physics and astrophysics.  But it is useful first to explore why the LIGO community was confident that this was a genuine event rather than noise.

The noise in the detector is large compared with the signal from realistic sources.  Even a comparatively strong event such as GW150914, which had a peak dimensionless strain of $h\approx 10^{-21}$ \cite{2016arXiv160203840T}, changes a 4~km distance by just $4\times 10^{-16}$~cm, or about 1/200 of the radius of a proton.  Thus passing trucks, the most minor earthquakes, electrical storms, etc., all move the mirrors by much more than they are moved by gravitational waves.  

Ground-based interferometers must thus be designed to reduce this noise as much as possible.  This takes many forms.  For example, mirrors are suspended via a sequence of four pendulums, each of which has a low resonant frequency, which therefore damps any ground motion at the frequencies of interest.  Seismometers around the sites monitor activity, which can then be incorporated into analyses.  Coincidence is required between the two detectors, within a light travel time (which is 0.01 seconds, given the $\sim$3000~km separation between Livingston, LA and Hanford, WA), and so on.  Yet even with great care, there is still significant noise.  Importantly, that noise is {\it not} Gaussian, because various non-Gaussian transients (glitches) occur \cite{2016arXiv160203844T}.

The non-Gaussianity of the noise means that the significance of a potential signal cannot be estimated just by dividing the strength of the signal by the standard deviation of the noise.  Instead, the LIGO team shifts the single-detector triggers from each of the two detectors and shifts them in time by amounts well in excess of the 0.01~second light travel time between the two sites \cite{2010PhRvD..82j2001A}.  This ensures that the coincidences after shifts cannot actually come from gravitational waves.  Doing this for many different shifts produces an estimate of the distribution of the amplitudes of non-Gaussian, transient glitches that are not captured by quasi-steady-state characterization of the detector's noise power spectral density.

As shown in Figure~4 of \cite{2016PhRvL.116f1102A}, all of the high-significance correlations produced using this shifting method included GW150914 in one of the data streams.  Doing the chance coincidence estimate this way therefore is extraordinarily conservative, and even by this estimate the time needed for coincident noise to equal the strength of the GW150914 event is much greater than 200,000 years \cite{2016PhRvL.116f1102A}.  A more representative estimate would remove GW150914 from the data streams.  When this is done, the significance becomes astronomical; a simple extrapolation of the signal to noise ratio (SNR) versus rate occurrence might suggest a chance probability less than $\sim 10^{-20}$ \cite{2016PhRvL.116f1102A}.  This was a real event.  It is worth noting that the LVT151012 event was also likely real (it had a chance occurrence rate of once per two years \cite{2016PhRvL.116f1102A}), and that three months of O1 data remain to be analyzed.  More detections are likely to be announced, if not from O1 data then from the next run (O2), which should probe a volume per time that is $\sim 2-3\times$ what was reachable in O1 for double neutron star binaries, and with a potentially larger factor of improvement for massive black hole binaries \cite{2016LRR....19....1A}.

The GW150914 event is fit extremely well using the general relativistic waveform from a double black hole binary with component masses $36^{+5}_{-4}~M_\odot$ and $29\pm 4~M_\odot$, and with low spin magnitudes \cite{2016arXiv160203840T}.  The ringdown after the merger was consistent with the expectations for black holes, although the SNR of the ringdown alone was too low to measure the multiple modes, including their damping time, that would be necessary to test the no-hair theorem \cite{1980ApJ...239..292D,2006PhRvD..73f4030B}.  The direction to GW150914 was only coarsely determined: the 90\% credible region spanned $\sim 600$ square degrees, or $\sim 1/70$ of the full sky \cite{2016arXiv160208492A}.  The addition of further detectors to the worldwide network (probably Virgo, then KAGRA, then LIGO-India among the $\gtorder 3$~km detectors) will be necessary to localize events better.  For strong events, localization to a few square degrees is anticipated \cite{2016LRR....19....1A}, which is still large by astronomical standards but is within the reach of many wide-field electromagnetic instruments at different wavelengths.

Following the discussion in \S~\ref{sec:gwbasics}, we assume that the binary is circular.  When the components of the binary are well separated and thus the binary evolves slowly, the gravitational wave signal in frequency space is narrowly distributed in frequency space at twice the orbital frequency (twice, because of the quadrupolar nature of gravity).  As the separation decreases, however, higher-order effects complicate the analysis.  These effects are often phrased in terms of a post-Newtonian (PN) expansion; that is, the acceleration of the bodies in a binary, or the gravitational waveform they emit, is presented as a series in $v/c$, where $v$ is a characteristic speed of motion of the binary.  Schematically, we have:
\begin{equation}
h_{+,X}=h_{+,X,{\rm Newt}}+(v/c)h_{+,X,{\rm 0.5PN}}+(v/c)^2h_{+,X,{\rm 1PN}}+\cdots
\end{equation}
where each coefficient $h$ is a function of the binary separation, the component masses, and the magnitude and direction of the individual spins.  The ``0.5PN" notation is a historical artifact that exists because $M/r$ was originally thought to be the correct expansion quantity, and $v/c\sim (M/r)^{1/2}$ in virial equilibrium.  See \cite{2014LRR....17....2B} for details about the coefficients.  

A particularly useful analytic way to represent the waveforms is the effective one-body (EOB) approach introduced by \cite{1999PhRvD..59h4006B}, in which the spacetime of a binary is treated as a deformed Kerr spacetime.  This approach overcomes the slow convergence of the standard PN series, and after calibration against the waveforms produced in full numerical general relativity it seems to be capable of representing faithfully the waveforms from even generic precessing binaries \cite{2016arXiv160302444H} as well as the waveforms from inspiraling neutron stars that are affected by static tides \cite{2015PhRvL.114p1103B} and possibly even dynamic tides (i.e., induced oscillations; see \cite{2016arXiv160200599H}).  The revolution in numerical relativity that began around 2004--2005 \cite{2004PhRvL..92u1101B,2005PhRvL..95l1101P,2006PhRvL..96k1102B,2006PhRvL..96k1101C} has been and continues to be critical to calibrate analytic approximations and to ensure that they are accurate even when the compact objects are close to merger.  The analytic approximations are necessary because the coalescence of lighter objects such as neutron stars will have thousands of cycles in the Advanced LIGO sensitivity band, and numerical simulations simply cannot run for that long within the time available for computations.  The approach taken by the LIGO Scientific Collaboration is to have a large set of templates that are matched against the data, which are dense enough in the relevant model parameters that they will have a strong overlap with any likely circular binary \cite{2015arXiv150802357U,2016arXiv160203839T,2016arXiv160404324M}.  Once a detection is made, more detailed Bayesian analysis is then used to obtain the best values and uncertainties for the masses, spins, direction and distance, binary orientation, and eccentricity.

Additional information could be obtained if there were an electromagnetic counterpart to the event; in principle, for example, the direction and redshift could be determined very precisely.  It is therefore of potential interest that the {\it Fermi} gamma-ray detector found a $\sim 1$-second transient that started 0.4~seconds after GW150914, and that was consistent with the (poorly localized) direction to GW150914 \cite{2016arXiv160203920C}.  The properties of the transient are similar to the properties of short gamma-ray bursts, which are believed to be produced by the merger of two neutron stars or a neutron star and a black hole (see \cite{2014ARA&A..52...43B} for a recent review of these events, and note that \cite{2016arXiv160204460L} point out that compared with other short gamma-ray bursts this event had a low luminosity and high spectral peak energy if it is associated with GW150914).  Indeed, had GW150914 involved the coalescence of a neutron star with either a low-mass black hole or another neutron star, the majority of the community would likely have accepted that the {\it Fermi} event was related to the gravitational wave event.  However, there is a strong expectation that mergers between two stellar-mass black holes will {\it not} produce detectable electromagnetic radiation, because the environments around such mergers are expected to have an extreme paucity of matter (see, e.g., the discussion section of \cite{2016ApJ...820L..36S}).  This, combined with the comparatively high probability of an unrelated signal of this strength during a plausible time window (a probability of $\sim 0.0022$ was quoted in \cite{2016arXiv160203920C}) and the lack of other reported electromagnetic or neutrino detections \cite{2016arXiv160208492A,2016arXiv160205411A,2016ApJ...820L..36S} suggests caution in identifying the events with each other.  In the meantime the electromagnetic follow-up community has certainly had its interest whetted!

\section{Implications for Gravitational Physics}
\label{sec:grav}

Precise tests of the predictions of GR existed before GW150914 (see \cite{2014LRR....17....4W} for a recent review), but none could directly probe gravity at spatial (high curvature) and temporal (rapid dynamics) scales anything like those near a black hole binary without involving potential astrophysical complexities.  Note also that electromagnetic tests of high curvature scales (e.g., \cite{2008LRR....11....9P,2016arXiv160207694J}), such as proposed searches for signatures of modifications of gravity in iron K$\alpha$ line profiles, or continuum X-ray spectra, or the shape of a black hole shadow, have typically not been performed using comprehensive parameter exploration.  That is, even if a modification of gravity changes (say) a line profile given that all other parameters are fixed, it needs to be demonstrated that {\it no} physically reasonable parameter combination in GR could produce the modified line profile (see, for example, some of the caveats listed at the end of \S~2 in \cite{2016CQGra..33e4001Y}).  Until such studies are done, it will not be clear to what extent electromagnetic observations can uniquely constrain effects due to modified gravity (see also \cite{2002A&A...396L..31A}).  

Many summaries of existing and prospective tests of gravity were written prior to the gravitational wave event; some recent examples that we recommend are \cite{2013LRR....16....7G,2015CQGra..32x3001B,2013LRR....16....9Y,2016CQGra..33e4001Y} .  There has also been detailed analysis of the implications of GW150914 itself for gravitational physics \cite{2016arXiv160203841T,2016arXiv160308955Y}.  Here we summarize some of the main points.  We draw in particular on \cite{2016arXiv160308955Y}, which has excellent discussions of many details that we omit for brevity.  We can, in some sense, divide tests into two categories: (1)~tests of the correctness of GR, and (2)~constraints on alternatives to black holes in GW150914, given the assumption that GR is correct.

\subsection{Tests of gravity with GW150914}

As emphasized by \cite{2016arXiv160308955Y}, there are multiple pillars of GR that could be tested by analysis of the gravitational waves from double black hole coalescences, including the strong equivalence principle, the no-hair theorem, the speed of gravity, the tensorial nature of gravity (i.e., without including scalar or vector components), and the four-dimensionality of spacetime.  A double black hole detection can not, of course, test deviations from GR that appear only in the presence of matter.  

A way to constrain deviations from GR is to use the parameterized post-Einsteinian (ppE) framework of \cite{2009PhRvD..80l2003Y}, in which the phase deviation from GR predictions is parameterized as
\begin{equation}
\delta\Phi_{\rm I,ppE}(f)={3\over{128}}(\pi {\cal M}f)^{-5/3}\sum_{i=0}^7\phi_i^{\rm ppE}(\pi {\cal M}f)^{i/3}
\end{equation}
where $\phi_i^{\rm ppE}$ are the ppE parameters.  The deformations to the GR waveform can then be considered one order (i.e., one value of $i$) at a time, in a theory-agnostic way, or individual alternate gravity theories can be mapped to their predicted values of $\phi_i^{\rm ppE}$.

In the LIGO-Virgo paper on tests of gravity, the authors varied the coefficients one by one while keeping the others fixed to their GR values \cite{2016arXiv160203841T}.  This exercise reveals that although at the 0PN (Newtonian) order binary pulsar observations still provide stronger constraints than did observation of GW150914, albeit in a different regime of gravity, at all higher orders GW150914 provides stronger upper limits on deviations from the predictions of GR.  These limits are not particularly strict: at the 90\% confidence level, the 0.5PN, 1PN, and 1.5PN terms could have fractional deviations of order unity from their GR values.  At higher orders (other than the 2.5PN order, which is degenerate with a uniform phase shift), the allowed fractional deviations are even greater.  Thus in this sense there is still considerable room for deviations from GR.  However, this room exists in large part because there is not yet a comprehensive framework to model deviations from GR in the late portion of the coalescence including the merger.  A different perspective on what deviations are allowed is that \cite{2016arXiv160203841T} find that after removing the best-fit waveform, less than 4\% of the signal remains.  There are, therefore, very restricted possibilities for generic deviations from GR.

The approach of \cite{2016arXiv160308955Y} is to divide deviations from GR into {\it generation mechanisms}, which they define as deviations that operate close to the binary, and {\it propagation mechanisms}, which they define as deviations that accumulate with distance traveled.  They emphasize that propagation mechanisms can typically be constrained far more tightly than generation mechanisms, because any process that leads to an accumulation of phase with distance (such as a frequency-dependent propagation speed for gravity, which would emerge if the graviton were massive) has an enormous amount of time to accumulate (roughly a billion years for GW150914) compared to the $\sim 0.2$~s duration of the event.

Indeed, when \cite{2016arXiv160308955Y} use this formalism to compare the details of GW150914 with the expectations from various specific theories, they find that the constraints on generation mechanisms are not strong, at least based on the coalescence up to the point of merger (as we indicate above, it is probable that the merger places significantly stronger constraints on generation mechanisms, but it is not possible at this time to determine the nature of those constraints).  For example, without prior knowledge of the spin magnitudes and component masses, they find that Einstein-dilaton Gauss-Bonnet \cite{1987NuPhB.293..385M}, dynamical Chern-Simons \cite{1974AoM..99..48,2009PhR...480....1A,2013PhRvD..87h4058Y}, and scalar-tensor theories (e.g., \cite{1999PhRvL..83.2699J}) cannot be constrained meaningfully because of degeneracies with the masses and spins, and Einstein-{\AE}ther theory \cite{2001PhRvD..64b4028J,2008arXiv0801.1547J} and ideas involving extra dimensions (\cite{2011PhRvD..83h4036Y}; see \cite{2009ApJ...705L.168G} for a limit based on the lack of Hawking evaporation of black holes in globular clusters) or time variation of Newton's constant $G$ \cite{2010PhRvD..81f4018Y} are better constrained by other measurements, particularly those of binary pulsars.  

In contrast, propagation mechanism constraints have been enhanced significantly by analysis of GW150914.  For example, if the dispersion measure of a massive graviton is assumed, GW150914 places an upper limit of $m_gc^2<2.2\times 10^{-22}$~eV on the mass $m_g$ of the graviton (\cite{2016arXiv160203841T}; see \cite{1998PhRvD..57.2061W} for the original proposal of this idea).  This is significantly better than the limits available from Solar System \cite{1988PhRvL..61.1159T} or binary pulsar \cite{2002PhRvD..65d4022F} tests, and is also better than the limits that can be obtained by the lack of gravitational Cerenkov slowing of high-energy cosmic rays in certain parameter regimes \cite{2015PhRvD..92j4036K}.  There are slightly stronger, but more model-dependent, limits on the graviton mass that can be placed by the existence of stellar-mass black holes \cite{2013PhRvD..88b3514B}.  The GW150914 limit on the mass of the graviton is significantly weaker than bounds from observations of galaxy clusters \cite{1973CaJPh..51..431H,1974PhRvD...9.1119G}, but those bounds also have potential systematic errors due to our lack of precise knowledge about the distribution of dark matter (and indeed massive gravity models on cosmological scales are often inconsistent with data).  These considerations apply straightforwardly to Fierz-Pauli theory \cite{1939RSPSA.173..211F} and Lorentz-violating massive gravity \cite{2004hep.th....7104R}, but theories with more complex dispersion relations (such as bigravity \cite{2012JHEP...02..126H}) are not as easy to assess.  

\subsection{Constraints on alternatives to black holes}

Suppose now that GR is correct.  Could objects other than black holes have merged to produce GW150914?  Significant constraints on such alternatives come from analysis of the ringdown portion of the waveform.  The short damping time of $\tau\approx 4$~ms \cite{2016arXiv160203840T}, the frequency of the dominant $l=m=2$ ringdown mode, and the lack of any other obvious modes, are as expected for black holes and are not easy to explain with alternate object classes.  

As an illustrative example of a class of objects that can be nearly ruled out using the GW150914 data, \cite{2016arXiv160208759C} explore the possibility that the merging objects formed a gravastar, which is an object that combines an exterior Schwarzschild spacetime with an interior de Sitter spacetime, with a layer of exotic matter in between \cite{2004PNAS..101.9545M,2007CQGra..24.4191C,2009PhRvD..80l4047P}.  It is important to recognize that even before the gravitational wave detection, such objects were not considered probable; there is no known mechanism to produce them, and the concatenation of spacetimes is arbitrary even if it is formally a solution to Einstein's equations.  The utility of the work by \cite{2016arXiv160208759C} is therefore to show that even an object whose existence is not physically motivated can be constrained by the GW150914 data.  Using the reasonable assumption that the angular momentum of the remnant is the same as it would be for a black hole, \cite{2016arXiv160208759C} find that the ringdown frequency and damping time of such an object are very distinct from what was seen in GW150914 unless the transition region between spacetimes is extremely close to the horizon.  Similarly, \cite{2015JPhCS.610a2044B} find that black holes with firewalls would have modified quasinormal mode frequencies, with fractional differences that are proportional to the mass of the firewall and amount to a few percent if the firewall mass is $\sim 10^{-4}$ of the black hole mass, although it is only at late times that the quasinormal modes become dominant.

More generally, \cite{2016arXiv160308955Y} note that the short damping time places large lower limits on the effective viscosity of any matter that makes up exotic black hole alternatives.  Boson stars, for example, would be expected to have damping times in excess of $\sim 100$~ms.  Modes with shorter damping times are less easy to constrain; for instance, \cite{2016arXiv160308955Y} find that modes with damping times less than $\sim 10$~ms, at any frequency, could have been undetected in GW150914 if their amplitudes were less than $\sim 30-100$\% of the amplitude of the dominant mode (although such a decrease in damping time would also increase the frequency of the ringdown).  Greater signal to noise ratios (SNRs) will be needed to produce stronger limits, and although this requirement may seem greedy given the SNR of 25 for GW150914 we note that full-sensitivity Advanced LIGO will find an SNR of $\sim 70$ for a similar event.  Data with SNR values of this order, or the combination of data from several similar events, are expected to reveal additional ringdown modes and will therefore test the no-hair theorem \cite{2007PhRvD..76j4044B}.  Even with just the data from GW150914, the constraints on exotic alternatives to black holes are extremely strong.

\section{Implications for black hole binary formation}
\label{sec:astro}

Assuming that GR is the correct theory of gravity, and that the objects that coalesced to produce GW150914 were black holes, what are the implications for the origin of the black holes?  For this discussion we take as our key pieces of information, in roughly decreasing order of importance, (1)~the masses of the black holes, (2)~the inferred rate of mergers, (3)~the limits on the spins of the holes, and (4)~the limit on the eccentricity of the orbit when it entered the Advanced LIGO band.  In more detail:

\begin{enumerate}

\item The masses of the two black holes were $29_{-4}^{+4}~M_\odot$ and $36_{-4}^{+5}~M_\odot$ \cite{2016arXiv160203840T}.  The high values of the masses, and their near-unity mass ratio, both pose meaningful constraints on formation scenarios \cite{2010ApJ...714.1217B,2010ApJ...715L.138B}.

\item The 90\% credible region for the rates of events such as GW150914 is $2-53$~Gpc$^{-3}$~yr$^{-1}$ \cite{2016arXiv160203842A}.  When less-certain events such as LVT151012 are included in the analysis, and some assumptions are made about the binary black hole mass distribution, the upper limit to the rate increases to $\sim 400$~Gpc$^{-3}$~yr$^{-1}$ \cite{2016arXiv160203842A}.  More detections will obviously narrow this range, and even GW150914 by itself eliminated the low end of the previous $0.1-300$~Gpc$^{-3}$~yr$^{-1}$ range of estimates \cite{2010CQGra..27q3001A}.

\item The dimensionless spin magnitude of the more massive black hole is less than 0.7 at 90\% credibility \cite{2016arXiv160203840T}.  Most of the constraint is on the component of the spin that is along the orbital axis; in-plane components have far less of an effect on the waveform.  The spin magnitude of the less massive black hole is essentially unconstrained.  From fits to the entire waveform, the spin magnitude of the final black hole is $0.67^{+0.05}_{-0.07}$ \cite{2016arXiv160203840T}.  This is consistent with the spin inferred from the merger and ringdown alone, but that inference by itself has much larger uncertainties.  Such a spin magnitude is consistent with the expected spin of the final black hole after a merger of two comparable-mass black holes with no spin \cite{2006PhRvD..73j4002B}.  The short duration of the signal meant that no precession was detected; precession could occur if at least one black hole had a significant spin that was misaligned with the orbital axis, which would add greatly to the information about an event \cite{2013PhRvD..87j4028G}.

\item There is no existing waveform model that consistently includes the effects of orbital eccentricity throughout the coalescence.  Rough estimates suggest that $e\ltorder 0.1$ at 10~Hz \cite{2016arXiv160203840T}.  

\end{enumerate}

An overall comment is that, although there are plentiful uncertainties, it appears likely that the black holes in GW150914 originated from gas that was significantly poorer than the Sun in its fraction of elements heavier than helium (in astronomical lingo, the gas was likely to be low-metallicity gas).  Indeed, the role of metallicity was highlighted several years before this event \cite{2010ApJ...715L.138B}. To understand this we will now have a brief digression on the formation of black holes via core-collapse supernovae (see \cite{2013RvMP...85..245B,2013CQGra..30x4002F} for recent reviews of the physics of these supernovae).

A star is supported against gravity by the radiation produced in the nuclear fusion reactions in its core.  Most of the nuclear binding energy is extracted in the fusion of hydrogen to helium, during the so-called ``main sequence" phase of stellar lifetimes.  For massive stars there can also be later fusion of helium to carbon and on up to the fusion of silicon to iron, which is short-lived both because little binding energy is available and because the core has to be extremely hot to overcome the large Coulomb barrier.  Iron has very nearly the greatest binding energy per nucleon of any nucleus at low pressure, which means that iron nuclei do not produce net energy with fusion.  Thus the standard, somewhat simplified, picture is that if the initial stellar mass is great enough, the fusion sequence is followed until inert iron is generated.  If the iron core exceeds the ``Chandrasekhar mass" \cite{1931ApJ....74...81C} $M_{\rm Ch}\sim 1.3-1.4~M_\odot$ that can be supported by degeneracy pressure gradients (the mass limit in real cores is increased somewhat by thermal pressure gradients; see \cite{2013RvMP...85..245B}), then it collapses on nearly a free-fall time of $\sim 1$~second.  If the collapse is halted by the formation of a neutron star then shocks can form and produce a supernova, although the required simulations are so complex that even after decades of work there are many details that are unclear.  The formation of black holes, rather than neutron stars, is even less clear, but presumably enough extra mass falls to the center that the total mass of the core exceeds the $\sim 2-3~M_\odot$ maximum mass of a neutron star.

Given this simplified description one might have guessed that stars that begin their lives with masses $m\gtorder M_{\rm Ch}$ end their lives as supernovae and that stars with masses above the neutron star maximum would finish as black holes, but the situation is more complicated \cite{2013CQGra..30x4002F}.  The point most relevant to our discussion of metallicity is that massive stars, which therefore have high luminosities, have very strong radiation-driven winds that can easily drive away tens of percent of the original mass of the star \cite{2000ARA&A..38..613K}.  The strength of these winds depends strongly on atomic line opacities; this might appear counter-intuitive, given that in the rest frame of an atom lines have significant opacity only over a very narrow range of photon energies, but broadening of lines by, e.g., thermal and collisional processes means that if there are enough lines they can overlap and produce a large average opacity.  Only heavy elements such as iron have enough electrons to produce many lines.  A reduction of the metallicity therefore reduces the opacity and thus the rate at which stellar mass is lost to winds.

Hence low-metallicity stars might form with somewhat higher average mass than solar-metallicity stars, and low-metallicity stars are likely to retain more of their original mass by the time their cores collapse.  It is thus expected that low-metallicity stars can produce significantly higher black hole masses than solar-metallicity stars.  The heaviest stellar-mass black hole known in our Galaxy is only $M\approx 15~M_\odot$ (see \cite{2011ApJ...741..103F}; note that previous estimates of the mass of the black hole in IC10 X-1 are unreliable \cite{2015MNRAS.452L..31L}), and the best current thinking is that to produce $\sim 30~M_\odot$ black holes the metallicity must be less than $\sim 10-50$\% of solar.  In particular, \cite{2010ApJ...715L.138B} showed that lower metallicities increase the likely masses of black holes and increase greatly the rate of gravitational wave detections.  Indeed, they predicted that the first event seen using Advanced LIGO would be a double black hole coalescence.  For recent papers discussing this point in the context of GW150914, see \cite{2016ApJ...818L..22A,2016arXiv160204531B}.

Even more massive black holes might evolve from stars.  \cite{2010MNRAS.408..731C} found three stars in the Large Magellanic Cloud (metallicity $\sim 0.5$ times solar) with current masses in excess of $150~M_\odot$.  If this means that stars can form up to hundreds of solar masses, then \cite{2014ApJ...789..120B} found that $\gtorder 100~M_\odot - 100~M_\odot$ black hole mergers might originate in massive stellar binaries.  Given that at design sensitivity Advanced LIGO will be able to see massive black hole binaries in a much larger volume than less-massive binaries (see Figure~6 of \cite{2014ApJ...789..120B}), there is the exciting prospect that a universe of massive black holes may open up \cite{2010ApJ...722.1197A}.

\begin{figure}
\centering
\hspace*{-0.5cm}\includegraphics[scale=0.5]{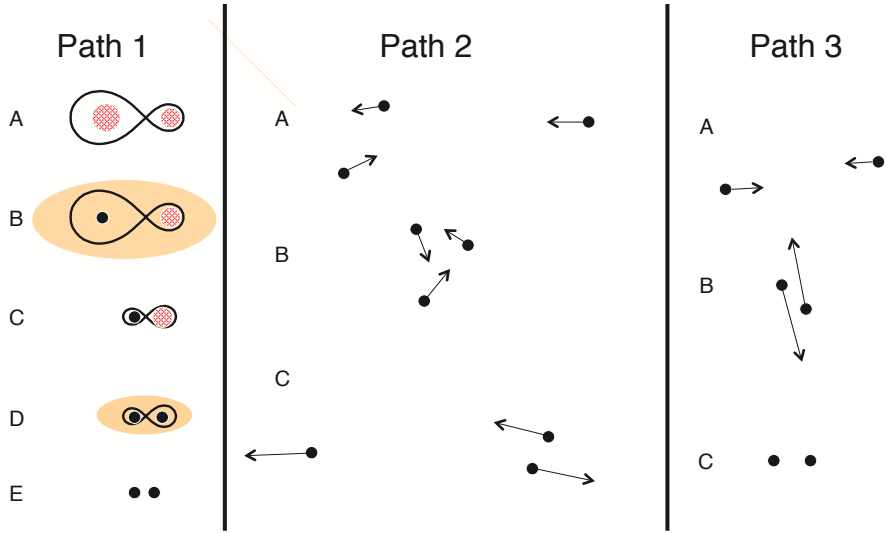}
\caption{
Schematic diagram of different ways to produce a merging stellar-mass black hole binary.  {\it Path 1}: Isolated binaries (see section \ref{sec:isolated}).  In panel A we see the initial system, which has two massive stars in a binary.  The figure-eight contour around the two stars represents the Roche surface, beyond which matter could flow to the other star.  In panel B, the more massive star has evolved to become a giant, so that it overflows its Roche lobe and the system enters a common envelope phase (represented by the light oval surrounding the system).  The second star, and the core of the first star, spiral closer via gas drag, which tightens the binary.  In panel C we see the now-tightened system after the first star has evolved to a black hole but the second star has not yet evolved.  In panel D the second star has expanded and the system is in a second common envelope phase, which leads ultimately to the even closer binary black hole in panel E.  There are many variants of this path; for example, if the stars are very strongly rotationally or tidally mixed, then they may not go through a giant phase, and this could lead to larger black hole masses.  {\it Path 2}: Dynamical formation (see section \ref{sec:dynamical}).  Here we envision a dense stellar system such as a globular cluster or the central region of a galaxy.  In panel A we see a binary at the left, with a third star approaching from the right.  In panel B the third star is interacting closely with the other two stars.  This three-body phase is formally chaotic, but the tendency is that the two most massive of the three stars form the ultimate binary, and that the final binary (at the right in panel C) is more bound than the initial binary.  Multiple interactions of this type can tighten a binary and thus speed up a merger.  {\it Path 3}: Gravitational two-body capture (see section \ref{sec:capture}).  If two initially unbound black holes (panel A) pass close enough to each other, then in their pericenter passage (panel B) they can radiate enough energy in gravitational waves to be bound into a tight and initially high-eccentricity binary (panel C).  There are other paths to merger that have been considered (via, e.g., the influence of the outer star in a triple system), but this is a representative set of possibilities.
}
\label{fig:paths}
\end{figure}

With these considerations in mind we now discuss different possible origins for the black holes in GW150914: as an isolated massive binary, as a binary in a dense stellar system, and as primordial black holes formed in the first second of the universe, before Big Bang nucleosynthesis.  We then comment on the implications for the stochastic background that would be produced by individually unresolved binaries of this class.  Figure~\ref{fig:paths} summarizes some of the main origins that have been considered. 

\subsection{Isolated binaries}
\label{sec:isolated}

One origin for a double black hole system simply involves having two massive stars in a dynamically isolated binary (i.e., a binary that never interacts with another star).  Indeed, the most massive stars we see almost always have a companion star (the binary fraction for massive stars is at least $\gtorder 80$\%; see \cite{2013ARA&A..51..269D}), and the secondary has a comparable mass to the primary (see \S~3.1.4 of \cite{2013ARA&A..51..269D} for a discussion of the dependence of the mass ratio distribution on the separation and mass of the binary).  If both stars are massive enough to evolve to black holes, there are still some obstacles to overcome before the system becomes a double black hole binary that can coalesce within the $\sim 10^{10}$~yr age of the universe.  For example, if the stars start too far apart, their evolution to black holes will leave them separated by such a distance that coalescence takes much longer than ten billion years.  If the stars start too close, they may merge before both black holes are formed.  In addition, the core collapse that forms a black hole can impart a kick of tens of km~s$^{-1}$ or more to the hole (see, e.g., \cite{2016ApJ...819..108B,2016MNRAS.456..578M} for recent discussions), which might unbind the binary.

These effects are incorporated in parameterized ways in various binary population synthesis codes (e.g., \cite{1996ApJ...466..234L,1998ApJ...496..333F,1998A&A...333..557D,2016ApJ...819..108B}; see \cite{2014LRR....17....3P} for a recent discussion of the evolution of compact binaries).  The simulators are in a difficult position because massive stars are rare, and because some of the key phases of the evolution are extremely tough to observe, which means that there is reliance on our often-incomplete theoretical understanding of the relevant processes.  One of those processes was mentioned above: supernova kicks.  The complexity of supernova simulations (and the very small kinetic energy of the kick compared to the overall supernova energetics) means that the distribution of kick speeds, and importantly the distribution of their direction relative to the orbital axis or the spin axis of the parent star, are highly uncertain.

Even more difficulties attend the so-called common envelope phase of binary evolution.  As an isolated star runs out of its current nuclear fuel (hydrogen, then helium, carbon, and so on), it transitions to a giant phase.  A star in the fully-developed giant phase has a dense core surrounded by a tenuous envelope.  The star's radius can easily increase by a factor of $100-1000$ during this phase.  A key part of the evolution of an isolated binary to a coalescing double black hole system is that if the stars are close enough that the giant's envelope overlaps the companion, the companion experiences gas drag and thus spirals inward, reducing the separation.  If the star has a smooth density gradient rather than a relatively sharp core-envelope structure, the companion could spiral all the way in and thus reduce BH-BH merger rates for solar metallicity systems \cite{2007ApJ...662..504B}.  If the companion is a neutron star or black hole, this might produce a ``Thorne-$\dot{\rm Z}$ytkow object" \cite{1975ApJ...199L..19T,1977ApJ...212..832T} in which the core consists of a compact object while the rest of the star looks like a relatively normal giant (although more investigation needs to be performed to determine whether such configurations are stable).  But to produce two black holes, the inspiral of the companion must eject the envelope so that the companion, now at a significantly reduced orbital radius, orbits the nearly-bare core.  This sequence can happen twice, and can lead to a double black hole system that is close enough that gravitational radiation will drive its merger in less than a few billion years.  The physics of the common envelope phase is tremendously complex, and even numerical simulations are not definitive; see \cite{2013A&ARv..21...59I} for a thorough summary of many of the issues.

Although the common envelope phase is considered essential in the standard picture of evolution from isolated massive binary to black hole binary merger, it has been pointed out that other processes could operate if the stars rotate rapidly, or if they have a strong tidal interaction with a close companion \cite{2016MNRAS.tmp..160M,2016A&A...588A..50M}.  The point is that the core-envelope structure develops because once a fused nucleus (say, helium on the main sequence, or carbon or heavier nuclei later on) is produced, it stays in the core and is not replaced by fresh fuel.  But rapid rotation can generate mixing throughout the volume of the star.  This would lead to a constant resupply of fuel.  \cite{2016MNRAS.tmp..160M,2016A&A...588A..50M} suggest that this could mean that there would never be a giant phase, which could in turn mean that the stars never come into contact with each other.  This might also lead to the formation of extra-massive black holes, because the cores that collapse would be large.  If the stars start out close to each other, they could potentially merge within a few billion years or less even if their orbits are not shrunk by common envelope interactions.  

Many of these mechanisms are still under active discussion, and GW150914 will undoubtedly stimulate more work.  A recent analysis in the light of GW150914 \cite{2016arXiv160204531B} concludes that in low-metallicity environments $\sim 30~M_\odot - 30~M_\odot$ coalescences are reasonable, and that in particular comparable-mass mergers are the norm.  Such low-metallicity regions are common enough, even in the current universe, that \cite{2016arXiv160204531B} find a rate per volume of 200~Gpc$^{-3}$~yr$^{-1}$ in their standard model.  

The usual picture is that if stellar precursors to black holes are close enough to influence each other tidally, and if any supernova kick is small, the rotational axes of the black holes will probably be fairly well aligned with the orbital axis.  If they are, then additional alignment can happen during the long merger process, from gravitational radiation alone (e.g., \cite{2004PhRvD..70l4020S,2015PhRvD..92f4016G}).  A common conclusion has therefore been that we expect black holes formed from isolated binaries to have aligned spins.  \cite{2016arXiv160204531B} adopt a note of greater caution about this conclusion, and suggest that misalignment is a real possibility if there are significant kicks (given that systems with large initial misalignments do not tend to align due to gravitational radiation; see \cite{2007ApJ...661L.147B}).  That would be unfortunate, given that spin alignment or misalignment has been suggested as one of the few ways that we could determine the origin of a black hole binary.  It will be interesting to see what additional studies conclude.

\subsection{Dynamical formation}
\label{sec:dynamical}

Additional processes are possible in dense stellar environments.  In the solar vicinity the number density of stars is $\sim 0.15$~pc$^{-3}$ \cite{1978SvA....22..186L}.  This is a low enough density that in our 4.6 billion year history no star has passed close enough to eject our outer planets.  In contrast, the center of a globular cluster can easily reach a density of $10^5$~pc$^{-3}$ \cite{2010arXiv1012.3224H} and near the center of a galaxy the density can exceed $10^9$~pc$^{-3}$ \cite{2009ApJ...697.1861A}.  Even at these densities it is rare for stars to collide, but binaries have an effective cross section comparable to the area of their orbit and thus binary-single and binary-binary interactions are common.

Such interactions are formally chaotic.  Nonetheless, numerical simulations over the past several decades (such as the early classic \cite{1975MNRAS.173..729H}) have revealed a number of trends that can often be understood using thermodynamic analogies.  Consider first the gravitational interaction between two single stars in a system such as a globular cluster or a star cluster near the nucleus of a galaxy.  If the stars have different masses, then because the speed of each will be determined by the potential of the cluster, the speeds will likely be comparable to each other.  Thus the more massive star will have greater energy.  The tendency towards equipartition of energy then suggests, correctly, that the usual outcome will be for the more massive star to give energy to the less massive star.  In a gravitational potential, loss of energy leads to sinking toward the center of the potential.  Thus these interactions tend to drive heavier stars such as black holes toward the center of gravitational potentials where the stellar density is larger.  The sinking time is the energy relaxation time
\begin{equation}
t_{\rm rlx}\approx {0.3\over{\ln\Lambda}}{\sigma^3\over{G^2m\rho}}
\end{equation}
\cite{1987degc.book.....S}, where $\ln\Lambda\sim 10$ is the Coulomb logarithm, $\sigma$ is the velocity dispersion, $m$ is the mass of the star or black hole that is being followed, and $\rho$ is the overall mass density of the surrounding stars (the average mass of the surrounding stars has only a minor impact on $t_{\rm rlx}$ if it is significantly less than $m$).  For $\sim 10~M_\odot$ black holes in globular clusters, $t_{\rm rlx}\sim{\rm few}\times 10^7$~yr is typical.  In galactic centers, inside the central region where the supermassive black hole dominates the potential, $t_{\rm rlx}\ltorder 10^{10}$~yr for galaxies smaller than the Milky Way.

Thus black holes sink in dense stellar systems.  When they do so, they can encounter binaries and interact dynamically with them.  ``Soft" binaries (which, roughly, have internal gravitational binding energies less than the typical kinetic energy of the single, or ``field", stars with which they interact) tend to widen after such interactions.  Again, we can think of this in terms of energy equipartition.  If a low-energy star in a binary interacts with a high-energy field star, then the field star will tend to transfer energy to the binary, which widens the binary.  The cross section of interaction thus increases, and thus the rate of interactions goes up.  This is a runaway process, which results in the separation of the binary.  Thus soft binaries soften and separate rapidly, and they are therefore usually not found in dense stellar systems.

``Hard" binaries are those with internal gravitational binding energies greater than the typical kinetic energies of field stars.  Following the same logic as above, binary-single interactions tend to harden hard binaries.  Their cross section thus decreases, which means that hardening is steady rather than being a runaway process.  Hardening of a binary makes its gravitational potential energy more negative.  That energy is transferred to the kinetic energy of the field star and the binary, which thus both receive a kick.  Their energy is then shared with the rest of the cluster, and therefore binaries serve as an energy source for the cluster.

When there is a close interaction between a single star and a hard binary, that interaction is often very complicated, requiring many binary dynamical times to resolve.  That is, multiple passes through the system are needed before one of the stars is finally ejected.  A tendency seen in numerical systems is that the final binary is usually composed of the two most massive of the original three stars.  This can be understood by an appeal to phase space: lighter objects can move faster, so for a given amount of energy in the ejected star, the lightest of the three objects can access the largest volume of momentum space.  For comparable-mass objects this is an approximate tendency, but if black holes and lighter stars are involved then it becomes very probable that the black holes end up in the binary.  The complexity of the three-body interactions means that the eccentricity of the final binary can have a wide range when the binary is well-separated, but it is still difficult to have a pericenter distance so small that evolution of the binary due to gravitational radiation leaves a detectable eccentricity at the $\gtorder 10$~Hz frequencies detectable from the ground \cite{2004ApJ...616..221G,2006ApJ...640..156G,2014ApJ...784...71S}.

Thus even if a black hole in a dense stellar system is formed as a single object, or if the supernova that created it kicks the hole out of its binary, as long as the hole is retained in the system it can sink, replace a star in a binary, and possibly pair with another black hole that swaps in later.  The result is that this is a channel completely independent of the isolated-binary channel discussed above.  For example, all the swapping and hardening can lead to a successful merger even if stellar envelopes never overlap.

The rate of mergers via this mechanism has been estimated by many authors.  A recent estimate suggests a rate per volume of $\sim 5$~Gpc$^{-3}$~yr$^{-1}$ in globular clusters without a massive central black hole, with typical total masses of $\sim 30-60~M_\odot$ \cite{2016arXiv160202444R}.  This is on the low end of, but compatible with, the wide range of rates currently inferred using GW150914.  Dynamical processes are chaotic, which means that there is no reason to expect alignment of spins.  Measurable eccentricities at $\gtorder 10$~Hz are theoretically possible via this route to merger, but they should be rare \cite{2006ApJ...640..156G,2014ApJ...784...71S}.

Another dynamical effect, which is finding increasingly wide application in fields such as exoplanet research, is the Kozai-Lidov effect \cite{1962AJ.....67..591K,1962P&SS....9..719L}.  This effect applies to hierarchical triple systems: systems in which there is an inner binary orbited at a much larger distance by a third object.  Such a situation is stable.  If the inner binary is inclined significantly compared with the orbit of the third object, then over many orbits of the binary and the third object the relative inclinations of the binary and third object change periodically.  Over such a Kozai cycle, the semimajor axis of the binary remains essentially constant while its eccentricity undergoes large swings.  If the maximum eccentricity is high enough, the pericenter separation can be small enough that gravitational radiation takes over (e.g., \cite{2002ApJ...576..894M,2002ApJ...578..775B}).  Thus the gravitational effect of the third object can induce the merger of the binary even if the original binary would not have merged in ten billion years.  Such a hierarchical triple is a fairly common outcome of binary-binary mergers (see \cite{1984MNRAS.207..115M} and subsequent papers), and these systems might lead on somewhat rare occasion to pericenters small enough that there is a measurable eccentricity in the Advanced LIGO band \cite{2003ApJ...598..419W}.

Another possibility that has recently been explored is that a black hole binary in a galactic central region could be considered to be in a triple system with the central supermassive black hole of the galaxy \cite{2012ApJ...757...27A,2015ApJ...799..118P,2016arXiv160404948V}.   There are large uncertainties in the rates, but if black hole binaries are common and if the perturbations to the orbit around the supermassive black hole give the binary many opportunities to be in a Kozai-favorable configuration, rates per volume as high as $\sim 100$~Gpc$^{-3}$~yr$^{-1}$ might be produced by this mechanism \cite{2016arXiv160404948V}.

As an overall comment, the reason that mechanisms that are specific to dense stellar systems often lead to rates lower than the rates from isolated binaries is that only a small fraction of stars are in systems such as globular clusters or nuclear star clusters.  For example, our Galaxy has $\sim 10^7$ total stars in globular clusters, and $\sim 10^7$ stars in its nuclear star cluster, compared with $\sim 10^{11}$ stars overall.  Thus processes in dense stellar systems must be far more efficient than processes in the field to compete in terms of rates.  It was suggested by \cite{2014ApJ...789..120B} that the efficiency of dynamical interactions might be combined with the shear numbers of the majority of stars by noting that most stars are born in stellar associations or open clusters, which have lower densities and shorter lifetimes than globular clusters but which do provide opportunities for binary-single interactions and Kozai resonances, among other dynamical possibilities.  Even moderate incidence of these types of interactions in the millions to hundreds of millions of year lifetimes of the associations and open clusters could significantly increase the probability that a massive black hole system formed from a massive binary will eventually merge due to gravitational radiation.

\subsection{Gravitational captures and primordial black holes}
\label{sec:capture}

In principle, two black holes that are initially unbound to each other could pass close enough that the gravitational radiation they emit causes them to be a binary (\cite{1989ApJ...343..725Q}; see \cite{2006ApJ...648..411K,2013ApJ...777..103T} in the context of Advanced LIGO).  For reasonable mutual speeds, this requires a very close passage: the formulae of \cite{1989ApJ...343..725Q} imply that the required pericenter distance is less than
\begin{equation}
r_{\rm p,max}\approx 6\times 10^9~{\rm cm}~\left(\mu\over{15~M_\odot}\right)^{2/7}\left(M\over{60~M_\odot}\right)^{5/7}\left(v\over{10~{\rm km~s}^{-1}}\right)^{-4/7}\; ,
\end{equation}
where we have scaled to a $30~M_\odot - 30~M_\odot$ binary in a globular cluster with a velocity dispersion of 10~km~s$^{-1}$.  Thus although this has been considered as a possibility, it is likely to be a rare process.  For example, \cite{2013ApJ...777..103T} estimates that the rate per volume will be only $\sim 0.005-0.02$~Gpc$^{-3}$~yr$^{-1}$.  However, the required closeness of the pericenter for capture means that possibly tens of percent of such captures will lead to binaries that are visibly eccentric in the Advanced LIGO band.  Thus clear eccentricities might point to this process.  A more promising avenue of capture of black holes by each other has recently been explored \cite{2016arXiv160203831B,2016arXiv160204226S}, in which stellar-mass black holes that are both in the accretion disk of an active galactic nucleus have their inspiral assisted by gas drag.  Such processes might also lead to significant growth of stellar-mass black holes before they accrete onto the central supermassive black hole \cite{2012MNRAS.425..460M,2016ApJ...819L..17B}.  Because gas drag should circularize the orbits, this mechanism is likely to lead to orbits that are nearly circular by the time they enter the Advanced LIGO frequency range.

Along the same lines, \cite{2016arXiv160300464B,2016arXiv160308338S} suggest that GW150914 might have been the result of two primordial black holes capturing each other.  Black holes would act dynamically as dark matter, and if they formed before $\sim 1$~s after the Big Bang the baryons that produced them would be locked up and thus there would not be a conflict with light element nucleosynthesis.  \cite{2016arXiv160300464B} argue that if $\sim 10-100~M_\odot$ black holes (whose number density is not strongly limited by microlensing; see \cite{2016arXiv160300464B}) comprise most to all dark matter then the plausible capture rate is in the remarkably broad range of $1.1\times 10^{-4}$ to 1400~Gpc$^{-3}$~yr$^{-1}$ depending on very uncertain details regarding clustering.  However, accretion onto these black holes would produce extra $z\gtorder 10$ ionization that would leave an imprint on the cosmic microwave background \cite{1979GReGr..10..633B,1981MNRAS.194..639C,1995ApJ...438...40G,2000ApJ...544...43M,2001ApJ...561..496M}; \cite{2007ApJ...662...53R,2008ApJ...680..829R} argue that this limits the fractional contribution to dark matter from such black holes to $\ltorder 10^{-5}$, which suggests that this is an improbable mechanism.  Consistent with this, \cite{2016arXiv160308338S} find that the merger rate estimate from GW150914 implies that at most a small fraction of dark matter can be in $\sim 30~M_\odot$ black holes.

\subsection{Stochastic backgrounds}

For every GW150914-like event that is easily detectable, we expect many events that are sub-threshold for Advanced LIGO and its sister detectors.  These individually undetectable events will combine to produce a stochastic background.  An extremely convenient framework for computing the strength of a stochastic signal was developed by \cite{2001astro.ph..8028P}.  For a set of circular binaries, \cite{2001astro.ph..8028P} finds that the energy density in gravitational waves divided by the critical energy density (i.e., the energy density that would make the geometry of the universe flat), in a logarithmic range around a gravitational wave frequency $f$, is
\begin{equation}
\Omega_{\rm GW}(f)={8\pi^{5/3}\over{9}}{1\over{c^2H_0^2}}(G{\cal M})^{5/3}f^{2/3}N_0\langle(1+z)^{-1/3}\rangle\; .
\end{equation}
Here $H_0$ is the Hubble constant, $N_0$ is the current comoving number density of remnants, and $z$ is the redshift; $\langle(1+z)^{-1/3}\rangle\approx 0.74-0.8$ for a wide range of merger histories \cite{2001astro.ph..8028P}.  If we estimate the number density of the remnants by taking the estimated rate density of GW150914-like events of $2-53$~Gpc$^{-3}$~yr$^{-1}$ and multiplying by $10^{10}$~yr, we get $\Omega_{\rm GW}=5\times 10^{-11} - 1.4\times 10^{-9}$ at $f=25$~Hz.  This can be compared with the estimate of $\Omega_{\rm GW}(25~{\rm Hz})=2\times 10^{-10}-3.8\times 10^{-9}$ from \cite{2016arXiv160203847T}, who use the results of population synthesis codes with different assumptions.  The increase compared with our simple estimate is because we are not including lower-mass black hole binaries and because star formation (and thus black hole formation) was significantly more common at redshifts $z\sim 1-2$ than it is now, and the prompt formation of black hole binaries and mergers means that the overall number of mergers is greater than we computed using the present-day rate.  This is an encouragingly high number.  If the number remains high after other detections are reported, we can expect to detect the stochastic background in a few years of runs using Advanced LIGO at its design sensitivity, although it will be challenging to extract information about the binary population from the stochastic signal \cite{2016arXiv160402513C}.

\section{Summary and Future Prospects}
\label{sec:summary}

GW150914 is the first drop in what will become a deluge of events discovered using gravitational wave detectors.  Advanced LIGO will perform its O2 run later in 2016 and its O3 run in 2017, and will then transition to a steady stream of runs with occasional upgrades.  The design sensitivity to double neutron star coalescences is nearly three times as great as it was in O1, which means that the volume probed will increase by more than twenty times.  The volume improvement will be much greater for events that are most easily seen at frequencies $\ltorder 30$~Hz, such as massive black hole mergers, because the sensitivity at such low frequencies will improve by a factor $\gtorder 10$ \cite{2016LRR....19....1A}.  More ground-based detectors will come on line soon: Advanced Virgo possibly by late 2016, KAGRA within a few years after that, and LIGO-India by 2025.  The development of this worldwide gravitational wave detector network will improve sensitivity, will greatly improve prospects for good localization and possibly the identification of electromagnetic counterparts, and will boost the fraction of time that multiple detectors are operational.  There is also the possibility that different detectors will be optimized for different frequencies, e.g., some could be designed to be most sensitive at low frequencies and some at high.  

Technologies such as the squeezing of light (see \cite{2015PhRvD..91f2005M} for a discussion of this and other technology improvements) could significantly improve the performance of the detectors, particularly at high frequencies.    In turn, this would improve the prospects that analysis of gravitational waveforms from NS-NS or NS-BH binaries will inform us about the properties of those stars.  In particular, if the radius of these stars as well as their masses can be inferred with precision and reliability, this will serve as a valuable input to nuclear physics theories (see \cite{2013arXiv1312.0029M,2016RvMP...88b1001W,2016EPJA...52...63M} for a discussion of the current systematics-dominated estimates of radii, as well as of the good prospects for measuring radii using X-ray observations with the NASA mission {\it NICER} \cite{2012SPIE.8443E..13G} and the planned European Space Agency (ESA) mission {\it LOFT} \cite{2012ExA....34..415F}).  Plans are also being made for third-generation detectors such as the proposed Einstein Telescope \cite{2010CQGra..27s4002P}, which would have 10-km underground arms and many other advances; this could reach ten times the sensitivity of Advanced LIGO and would thus greatly increase both the rates and precision of observations.

There are also exciting possibilities for the detection of gravitational waves in other frequency bands.  At very low frequencies $\sim 10^{-17}-10^{-16}$~Hz, B-mode polarization in the cosmic microwave background would be a signal of primordial gravitational waves \cite{1997PhRvL..78.2054S,1997PhRvL..78.2058K}.  At higher frequencies $\sim 10^{-9}-10^{-6}$~Hz, pulsar timing arrays are improving their sensitivity; their first detection is likely to be of the stochastic background formed by a large set of supermassive black hole binaries, but there is some prospect for individual detections \cite{2015MNRAS.451.2417R}.  At yet higher frequencies $\sim 10^{-4}-10^{-1}$~Hz, eLISA is an ESA mission planned for launch in 2034 \cite{2013arXiv1305.5720C} that will involve three freely-falling test masses some millions of kilometers apart.  This will be sensitive to sources including gravitational waves from merging massive black hole binaries; intermediate-mass black holes, stellar-mass black holes and neutron stars spiraling into supermassive black holes; close white dwarf binaries that might be the precursors to Type Ia supernovae; and phase transitions from the early universe, and might serve as advance warning of GW150914-like mergers that would enter the ground-based frequency band within some years after the eLISA detection \cite{2013arXiv1305.5720C,2016arXiv160206951S}.  Combining all such experiments, gravitational wave astronomy will therefore span twenty orders of magnitude in frequency, which is similar to the range in electromagnetic waves and which will probe a vast set of phenomena.

In summary, GW150914 by itself has many exciting implications for gravity, exotic compact objects, and the formation and evolution of black hole binaries.  The best, however, is still to come.

\begin{acknowledgements}
We thank John Baker, Imre Bartos, Chris Belczynski, Emanuele Berti, Jordan Camp, Frans Pretorius, Kent Yagi, and Nico Yunes for many helpful comments on a previous version of this manuscript, the referees for extremely constructive reports, and the LIGO team for making the discovery.
\end{acknowledgements}

\bibliography{ms}

\end{document}